\documentclass[a4paper,11pt]{article}
\usepackage[utf8]{inputenc} 
\usepackage{jcappub} 

\usepackage[T1]{fontenc} 

\usepackage{commands}

\title{Dark energy loopholes some time after GW170817}

\author[a,1]{Lorenzo Bordin,}
\author[a,2]{Edmund J. Copeland,}
\author[a,3]{Antonio Padilla}

\affiliation[a]{School of Physics and Astronomy, University of Nottingham, \\Nottingham NG7 2RD, UK}

\emailAdd{lorenzo.bordin@nottingham.ac.uk}
\emailAdd{ed.copeland@nottingham.ac.uk}
\emailAdd{antonio.padilla@nottingham.ac.uk}

\abstract{We revisit the constraints on scalar tensor theories of modified gravity following the purge of GW170817. We pay particular attention to dynamical loopholes where the anomalous speed of propagation of the gravitational wave can vanish on-shell, when we impose the dynamical  field equations.  By working in the effective field theory formalism we are able to improve on previous analyses, scanning a much wider class of theories, including Beyond Horndeski and DHOST. Furthermore, the formalism is well adapted to consider the effect of inhomogeneous perturbations, where we explicitly take into account the fact that the galactic overdensities are pressureless to leading order.}

\begin{document}
\maketitle
\flushbottom

\section{Introduction}
Multi-messenger astronomy involving coordinated signals of electromagnetic radiation and gravitational waves provide a powerful tool for constraining fundamental theories of gravity. This was demonstrated emphatically by the merger of two neutrons stars at redshift $z \sim 0.01$, detected through a gravitational wave GW170817 and a burst of gamma rays GRB170817A \cite{GW170817, GRB170817A, Integral, combo, multi-mess}.   These effectively simultaneous twin observations constrained the speed of the gravitational wave through the cosmological medium to be identical to the speed of light to an accuracy of one part in a quadrillion!  This immediately led to a slew of papers  (see,  for example, \cite{Creminelli:2017sry, Sakstein, Ezquiaga:2017ekz, Baker, Amendola, Langlois, Crisostomi, Babichev:2017lmw, Kase, Copeland:2018yuh}) examining the implications for modified theories of gravity, especially those that are designed to reproduce the effects of dark energy through their long range modifications (for relevant reviews of modified gravity, see \cite{tonyrev,joycerev}).  The multi-messenger probe has proven particularly adept at constraining scalar tensor theories, such as Horndeski \cite{Horndeski, Horn2} or Beyond Horndeski \cite{Bhorn, Bhorn2},  where the gravitational wave will generically propagate through the cosmological background at a different speed to its electromagnetic counterpart. This happens even though the gravitational wave travels through regions of higher density where screening mechanisms are expected to operate \cite{Jimenez,Lombriser1,Lombriser2}. 

Modified gravity took a hit after LIGO/Virgo announced this result.  Large classes of scalar-tensor theory were declared incompatible with the data from the neutron star event \cite{Creminelli:2017sry, Sakstein, Ezquiaga:2017ekz, Baker}, or else irrelevant to gravity at  sufficiently large distances.  It is important to realise that these papers did not declare all modified gravity models incompatible. They still left room for scalars conformally coupled to curvature and some models with derivative couplings, such as  Kinetic Gravity Braiding \cite{KGB}. Furthermore, subsequent investigations identified possible loopholes in the original analyses.  For example, in \cite{claudia}, it was noted that the frequency of the gravitational wave was very close to the effective field theory cut-off of the relevant dark energy models, raising doubts as to whether or not we are entitled to constrain these theories without further knowledge of their ultra-violet corrections.  In \cite{Copeland:2018yuh}, a dynamical loophole was considered. It was shown that the anomalous speed of propagation did not need to vanish identically, rather it could vanish {\it on-shell}, on account of the scalar equation of motion.  At the level of the homogeneous background, a candidate theory from the Horndeski class was shown to exploit exactly this behaviour. A preliminary analysis suggested that this loophole would not survive the consideration of inhomogeneous cosmological perturbations, although the analysis did not exploit all of the dynamical data. 

In this paper, we revisit the idea of dynamical loopholes using the effective theory of dark energy  \cite{Cheung:2007st, Gubitosi:2012hu}. This has the advantage that we can investigate the existence of loopholes in a much broader class of models including Beyond Horndeski\footnote{\label{prem} Beyond Horndeski loopholes were briefly considered in  \cite{Copeland:2018yuh} but were prematurely ruled out  using constraints coming from the decay of the gravitational wave,  drawn directly from  \cite{Creminelli:2018xsv}. However, these constraints only apply to the residual class of theories left after the analyses of  \cite{Creminelli:2017sry, Sakstein, Ezquiaga:2017ekz, Baker} and {\it not} the dynamical loopholes.} \cite{Bhorn, Bhorn2} and so-called DHOST models \cite{Langlois:2015cwa, DHOST2, Crisostomi:2019yfo}.  It also lends itself to a careful study of linearised perturbations, both homogeneous and inhomogeneous.  Previous analyses did not consider the nature of the inhomogeneous source.  Here we assume it is non-relativistic, consistent  with galactic overdensities.  This means we have a vanishing inhomogeneity in the pressure perturbation, and another dynamical zero in the metric equations of motion,  in addition to the scalar equation of motion.

With all of these new ingredients, we perform a complete analysis of the full class of theories from Horndeski to Beyond Horndeski and DHOST. At the level of a homogeneous background, we recover the loophole theory found in  \cite{Copeland:2018yuh}, along with some generalisations. We then consider this extended class of loopholes in the presence of inhomogeneous perturbations, once again using dynamical knowledge of a vanishing scalar equation of motion (although now including inhomogeneities) and also the vanishing pressure perturbation. At the risk of ruining the punchline, we refer the reader to the conclusions for the final outcome of this analysis.

The rest of the paper is organised as follows. In section \ref{sec:EFT_action} we review the effective field theory (EFT) approach to dark energy, following  \cite{Cheung:2007st, Gubitosi:2012hu}. We review the generic approach of \cite{Creminelli:2017sry, Sakstein, Ezquiaga:2017ekz, Baker} in section \ref{sec:Paolos_argument} before describing the loophole found in \cite{Copeland:2018yuh}, along with generalisations,  in section \ref{sec:loophole}. In section \ref{sec:perts} we study the impact of inhomogeneities,  fully taking into account the pressureless nature of the inhomogeneous perturbations.  Finally, in section \ref{sec:conc}, we conclude.

\section{EFT of dark energy}
\label{sec:EFT_action}

The current expansion of the universe can be explained assuming the existence of a scalar field $\phi$ whose energy density fills the universe. 
The time-dependent homogeneous solution $\bar \phi(t)$ generates a preferred foliation that slices the spacetime into a Friedmann Robertson Walker  (FRW) metric. 
To work out the predictions for large scale structure  surveys, we study the perturbations of $\phi$ around a background that non-linearly realises time diffeomorphism invariance (time diffs).
Instead of focussing on a particular theory of dark energy, we use an effective approach, writing down the most general Lagrangian for $\phi$ and then expanding it around its time dependent background. 
It is therefore more convenient to develop a geometrical approach to expand and categorise perturbations rather than expand the scalar field $\phi =  \bar \phi+\delta \phi$.
In the following we outline this procedure, following references \cite{Cheung:2007st, Gubitosi:2012hu}.

\subsection{Dark energy action in unitary gauge}

We set the gauge by choosing the time coordinate to be a function of $\phi$, $t = t(\phi)$, in such a way that the dark energy  field sits at its unperturbed value everywhere.
This is the so called \emph{unitary gauge}, in which the perturbations of the field are eaten by the metric.
The constant-time slices generated by $\phi$ foliate the spacetime, making the action for perturbations no longer invariant under time diffs.
It follows that, beside the genuinely 4-d covariant terms such as the Ricci scalar, one can also include objects which are constructed from the foliation.
For instance, now we can separately  consider the projection of tensors along the orthogonal and parallel directions to the surface.
This is done by contracting tensors with the normal vector $n_\mu$ or with $h_{\mu\nu}$ the 3-metric, respectively given by
\be
n_\mu = \frac{\d_\mu \phi}{\sqrt{-\d_\nu\phi \d^\nu\phi}} \qquad \mbox{and} \qquad h_{\mu\nu} = g_{\mu\nu} + n_\mu n_\nu \,. 
\ee
All geometrical objects built from the foliation can be defined using these two projectors. 
Examples are the extrinsic curvature tensor $K_{\mu\nu} = h^\rho_\mu h^\sigma_\nu \nabla_\rho n_\sigma$ or the intrinsic curvature of the 3-dimensional surfaces ${}^{(3)}R_{\mu\nu\rho\sigma}[h]$.
The most general effective action is constructed by writing down all possible operators that are compatible with the remaining symmetries. 
The reduced symmetry pattern of the system allows many terms in the action. 
They can be categorised as follows:
\begin{itemize}
\item[i.] Terms which are invariant under all diffs: these are just polynomials of the 4-dimensional  Riemann tensor $R_{\mu\nu\rho\sigma}$ and of its covariant derivatives $\nabla_\mu$, contracted to give a scalar.
\item[ii.] Terms contracted with $n_\mu$. 
             Since in unitary gauge $n_\mu \propto \delta^0_\mu$, in every tensor we can leave free upper $0$ indexes. 
             For instance, we can use the $00$ component of the metric or of the Ricci tensor. 
             It is easy to check that these are scalars under spatial diffs.
\item[iii.] Terms derived from the foliation.
              This includes terms like $\mathcal D_\mu$, the derivatives of the induced 3-metric or the Riemann tensor ${}^{(3)}R_{\mu\nu\rho\sigma}$ that characterises the 3-d slices intrinsically; but one can use also objects that tell us how the hyper-surfaces are embedded in the 4-dimensional spacetime. 
              These are the extrinsic curvature $K_{\mu\nu}$ and the acceleration vector $A_\mu = n_\rho \nabla^\rho n_\mu$. The action contains all the possible scalars made by contracting these quantities.
\item[iv.] Since time diffs are broken all the couplings in front of the operators can be functions of time.
\end{itemize}
The most general action constructed using these ingredients takes the form
\be
S = \int d^4x \ \sqrt{-g} \ \mathcal L [g_{\mu\nu}, g^{00}, R_{\mu\nu\rho\sigma}, K_{\mu\nu}, \nabla_\mu, t] \,,
\ee
where the contractions are done with the metric $g_{\mu\nu}$, using the 3-metric does not lead to new interactions. 
Since the above action contains an infinite number of operators, we organise them in a derivative expansion: at lowest order in derivatives acting on the metric there are only polynomials in $g^{00}$, at first order we can use the trace of the extrinsic curvature $K$. 
The leading effective action, up to second order in derivatives is
\bea\label{eq:DE_action}
S_{EFT} &=& \int d^4 x \sqrt{-g} \Big[ \frac{M_*^2}{2} f  R - \Lambda  -c  g^{00}+ \nonumber\\
&& + \frac{m_2^4}{2} (\delta g^{00})^2 -\frac{m_3^3}{2} \delta K \delta g^{00} - m_4^2 \delta \K_2+\frac{\widetilde m_4^2}{2}\delta g^{00} \ {}^{(3)}R \nonumber\\
&&-\frac{m_5^2}{2} \delta g^{00} \delta \K_2 -\frac{m_6}{3} \delta \K_3 -{\widetilde m_6} \delta g^{00} \delta \G_2\,.
\eea
with 
\bea
&& \delta \K_2 \equiv \delta K^2 -\delta K_\mu^\nu \delta K_\nu^\mu\,, \qquad  \delta \G_2 \equiv \delta K_\mu^\nu {}^{(3)}R_\nu^\mu-\delta K {}^{(3)}R/2\,, \\
&&\delta \K_3 \equiv \delta K^3 -3 \delta K \delta K_\mu^\nu \delta K_\nu^\mu + 2 \delta K_\mu^\nu \delta K_\rho^\mu \delta K^\rho_\nu\,. \nonumber
\eea
In the above action,  $\delta g^{00}\equiv 1+g^{00}$ , $\delta K^\mu_{\ \nu} \equiv K^\mu_{\ \nu} - H h^\mu_{\ \nu}$, with $H \equiv \dot a/a$ being the Hubble rate, and $\delta K$ its trace.
While $M_*^2$ is a constant, the other parameters are slowly-varying time-dependent functions.
The first line of eq.~\eqref{eq:DE_action} consists of all the operators that start at the background level. 
The Friedmann equation and requirement that the dark energy stress-energy tensor is covariantly conserved force
the parameters $f $, $\Lambda $ and $c $ to satisfy
\be
\label{eq:fried_eq} 3 H^2 M_*^2 f - \Lambda - c = \rho_m\,, 
\ee
\be
\label{eq:conserved_stress_energy} 6Hc + \dot c + \dot \Lambda - 3 M_*^2 \dot f (2 H^2 + \dot H) = 0\,,
\ee
where $\rho_m$ is the matter density. Note that if $f=$ constant and $c=0$ the background dynamics is trivially equivalent the $\Lambda$CDM model and dark energy is driven by a cosmological constant, as opposed to some large distance modification of gravity. To properly explore modified gravity we really ought to  deviate from this trivial case.

In the second line of eq.~\eqref{eq:DE_action} there are those operators that start at quadratic order in perturbations, while
the last line contains cubic order operators. We did not attempt to write all possible cubic operators - only those that affect the graviton propagation speed. The action \eqref{eq:DE_action} describes both Horndeski and beyond-Horndeski theories. In principle. the parameters $m_i$ and  $\widetilde m_i$  are totally unconstrained  save for the fact that the action must be real (in particular, positive powers of $m_i$ and $\widetilde m_i$ can have either sign). Later on we will consider corrections that correspond to so called DHOST theories \cite{Langlois:2015cwa, DHOST2, Crisostomi:2019yfo}. 

\section{Dark energy immediately after GW170817}
\label{sec:Paolos_argument}

The action \eqref{eq:DE_action} describes the perturbations of the dark energy field and also gravity. Therefore, it captures  the motion of gravitational waves (GWs) travelling across the universe. These  may be affected by the presence of the time-dependent foliation that breaks diffs  (and hence Lorentz invariance on smaller scales) by changing the speed of propagation or even allowing them to decay into dark energy excitations \cite{Creminelli:2018xsv}: the situation is analogous to light travelling in a medium. 
Recently, the twin observation of GWs (GW170817) and its electromagnetic counterpart (GRB170817A ) coming from  a neutron star merger has put severe constraints on the speed of propagation of GWs relative to light  $c_T  /c_\text{light}=1 \pm \mathcal O(1) \times 10^{-15}$.  Note that in General Relativity, $c_{T, GR}  /c_\text{light}=1$ although generic modifications yield a deviation of this result. Therefore, this event has been used in \cite{Creminelli:2017sry, Sakstein, Ezquiaga:2017ekz, Baker}  to rule out many of the operators of the dark energy action \eqref{eq:DE_action}. 

Let us explore the argument of \cite{Creminelli:2017sry} in detail.
We start by expanding the action in scalar and tensor perturbations.
To expand the metric we use the ADM decomposition, so the line element is parametrised as $ds^2 = -N^2 dt^2 + h_{ij} (dx^i + N^i dt)(dx^j + N^j dt)$, and we work in the Newtonian Gauge, defined by
\be 
N^2 = 1 + 2\Phi\,, \qquad N_i = 0 \,, \qquad h_{ij} = a^2 (1-2\Psi) (e^\gamma)_{ij}\,,
\ee
with $\d_i\gamma_{ij} = \gamma_{ii} = 0$.
Time diffs are  restored by defining the Goldstone boson $\pi(x, t)$. 
Since we are interested in studying the dynamics of GWs,  we focus on the part of the action that is at least quadratic in the graviton perturbations.
We also keep cubic operators that contain two gravitons and one scalar perturbation.
This is because scalar perturbations with very long wavelength are seen by the astrophysical GWs (whose wavelength is $\sim 10^3$ km) as a local change of the FRW background, and the value of the couplings in the dark energy action depend on the particular background for the effective theory.
Therefore, an astrophysical GW travelling in toward us experiences many different FRW histories.
The full dark energy action, expanded at second order in $\gamma_{ij}$ and up to first order in scalar perturbations is given by eq.\eqref{eq:full_eft_action}. Here we just report those terms that contribute to $c_T$:
\be \label{eq:graviton_quad_action}
S = \frac 18 a^3 \[ A \dot \gamma_{ij}^2 - a^{-2} B \(\d_k \gamma_{ij} \)^2 + \dots\]\,,
\ee
with
\bea \label{eq:c_T}
&& A \equiv M_*^2 f \( 1 -\Phi - 3\Psi + \dot f / f \, \pi \) + 2 m_4^2 + 2 m_5^2 \( \Phi - \dot \pi \) - 2 m_6 \( H \Phi + \dot H \pi + \dot \Psi + \d^2\pi/a^2 \)\,, \\
&& B \equiv M_*^2 f \( 1 +\Phi - \Psi + \dot f / f \, \pi \) + 2 \widetilde m_4^2 \(\Phi - \dot\pi\) +2 \widetilde m_6 \( \dot \Phi - \ddot \pi\) + \dot{ \widetilde m}_6 \( \Phi -\dot \pi\) \,.
\eea
To compare this result with the predictions of General Relativity, it is useful to define the parameter 
\bea \label{eq:alpha_t}
\alpha_T &\equiv& \frac{c_T^2 - c_{T, GR}^2}{c_{T, GR}^2} \nonumber \\
&=& -2 m_4^2 \ \frac{2m_4^2 + M_*^2 f \( 1 + \Phi + 3\Psi -\dot f / f \pi \)}{\( M_*^2 f + 2m_4^2 \)^2} + 2 \widetilde m_4^2 \ \frac{ \Phi - \dot \pi }{ M_*^2 f + 2m_4^2 } - 2 m_5^2 \ \frac{ M_*^2 f (\Phi -\dot \pi)}{\( M_*^2 f + 2m_4^2 \)^2} \nonumber \\
&& + 2 m_6 \ \frac{M^2_* f \( H\Phi + \dot H \pi +\dot \Psi+ { \d^2\pi/a^2} \)}{\( M_*^2 f + 2m_4^2 \)^2} +  2 \widetilde m_6 \ \frac{\dot \Phi - \ddot \pi}{M_*^2 f + 2m_4^2} + \dot {\widetilde m}_6 \ \frac{\Phi -\dot \pi}{M_*^2 f + 2m_4^2}  \,,
\eea
where the last equality  holds up to first order in perturbations.  We see immediately that at background level the requirement $\alpha_T = 0$ is achieved by setting $m_4 = 0$.
However, to robustly set $c_T$ to coincide with the prediction of GR, we should also set to zero those couplings whose operators are turned on by a long scalar perturbation, i.e.~we set $\widetilde m_4^2 = m_5^2$ and  $m_6 = \widetilde m_6 = 0$.  This greatly reduces the phase space of the available dark energy theories. 

Similar results can be derived if we work in the covariant formalism \cite{Baker}.
After having restored the scalar field $\phi$, the Horndeski and Beyond-Horndeski actions are given by
\be \label{eq:B-H_action}
S =  \int d^4x \sqrt{-g} \sum_{n=2}^5 \mathcal L_n [\phi,  X]\,,
\ee 
where $X \equiv \d_\mu \phi \d^\mu\phi$ and
\bea
\mathcal L_2 &=& G_2(\phi ,X)\,, \\
\mathcal L_3 &=& G_3 (\phi, X) \Box \phi \,, \\
\mathcal L_4 &=& G_4 (\phi , X) R - 2 G_{4,X} \nabla_{[ \mu_1} \nabla^{\mu_1} \phi \nabla_{\mu_2 ]} \nabla^{\mu_2} \phi \nonumber \\
&& \qquad  + F_4(\phi, X) {\epsilon^{\mu\nu\rho}}_\sigma \epsilon^{\mu'\nu'\rho'\sigma} \nabla_\mu \phi \nabla_{\mu'} \phi \nabla_\nu \nabla_{\nu'} \phi  \nabla_\rho \nabla_{\rho'} \phi \,, \\
\mathcal  L_5 &=& G_5 (\phi, X) G_{\mu \nu} \nabla^\mu \nabla^\nu \phi + \frac{G_{5,X}}{3} \nabla_{[ \mu_1} \nabla^{\mu_1} \phi \nabla_{\mu_2} \nabla^{\mu_2} \phi \nabla_{\mu_3 ]} \nabla^{\mu_3} \phi \nonumber \\
&& \qquad + F_5(\phi, X) \epsilon^{\mu\nu\rho\sigma} \epsilon^{\mu'\nu'\rho'\sigma'} \nabla_\mu \phi \nabla_{\mu'} \phi \nabla_\nu \nabla_{\nu'} \phi  \nabla_\rho \nabla_{\rho'} \phi \nabla_\sigma \nabla_{\sigma'} \phi \, .
\eea
In this formalism the quadratic action for the graviton on a homogeneous background is given by  \cite{kob}
\be
S^{(2)}_T = \frac{1}{8} \int d^4x \, a^3 \[\mathcal G_T \dot \gamma_{ij}^2 -\frac{\mathcal F_T}{a^2} \(\d_k \gamma_{ij}\)^2\] \,,
\ee
with
\bea
\mathcal F_T &=& 2 G_4 + X G_{5,\phi} - 2X\ddot \phi G_{5,X} \,, \\
\mathcal G_T &=& 2G_4 - 4X G_{4,X} - X G_{5,\phi} + 2X^2 F_4 - 2H X \dot \phi (G_{5,X}+3X F_5).
\eea
Therefore the parameter $\alpha_T$ can be expressed as
\be \label{eq:alpha_t_covariant}
\mathcal G_T \ \alpha_T = -2 X \ddot\phi G_{5,X}+4XG_{4,X}+2XG_{5,\phi} - 2X^2 F_4 +2HX\sqrt{-X}(G_{5,X}+3XF_5)\,.
\ee
Requiring  $\alpha_T = 0$ for any background means that it should vanish independently of the values of $H$, $\dot H$, $\dot\phi$ and $\ddot \phi$.  
This implies $G_{5,X} = 0$, $F_5 = 0$ and $2 G_{4,X} - XF_4 + G_{5,\phi} =0$. 
Therefore, $G_5$ can only be a function of $\phi$, the Beyond Horndeski term $F_5$ must be absent and $F_4$ is fixed in terms of the derivatives of $G_4$ and $G_5$.

\section{Looking for a dynamical loophole}
\label{sec:loophole}

Recently, Ref.~\cite{Copeland:2018yuh} uncovered a loophole in the argument of the previous section that could potentially rescue an entire class of theories that had been previously discarded. The idea was that gravitational waves must travel at the speed of light only in physical systems that satisfy the classical equations of motion.  The authors of \cite{Copeland:2018yuh} worked in the covariant formalism and used the \emph{homogeneous} scalar equation of motion to express $\ddot \phi$ in terms of $\dot \phi$ and $\dot H$, so that they could substitute it in eq.~\eqref{eq:alpha_t_covariant}.
This opened up a new region in the parameters space of potentially viable theories.

We begin by reviewing the loophole identified in \cite{Copeland:2018yuh}, To this end, consider a theory described by the action \eqref{eq:B-H_action}, with potentials given by 
\bea
  G_2 = -3 \mu W'''(\phi) X \sqrt{-X} + \Lambda - \frac{\nu e^W(\phi)} {X} \,, &\qquad &G_3 = -6 \mu W''(\phi) \sqrt{-X} \,, \\
  G_4 = \kappa_G + \frac 32 \mu W'(\phi) \sqrt{-X}\,, &\qquad & G_5 = - \frac{6 \mu }{\sqrt{-X}} \,.
\eea
Here, $\kappa_G$, $\mu$, $\Lambda$ and $\nu$ are constants, and the prime denotes a derivative w.r.t.~$\phi$.
Using eq.~\eqref{eq:alpha_t_covariant} one can verify that
\be 
\alpha_T  = \frac{\mu \, X^2}{2 \nu \dot \phi e^W (3H\mu -\kappa_G) } \varepsilon_\phi \,,
\ee
where $\varepsilon_\phi$ is the homogeneous scalar equation of motion, taken to vanish on-shell
\be \label{eq:EoM_rescued}
\varepsilon_\phi \equiv \frac{3 \nu e^W \( 2\ddot \phi - 2H \dot \phi + X W' \)}{X^2} = 0 \,.
\ee
This means the gravitational wave constraints are satisfied dynamically, at least for homogeneous configurations, thereby evading some of the conclusions drawn in \cite{Creminelli:2017sry, Sakstein, Ezquiaga:2017ekz, Baker}. However, the preliminary analysis of  \cite{Copeland:2018yuh} also suggested that this theory would not survive the necessary constraints in the presence of 
inhomogeneous perturbations. But the analysis did not make use of the available dynamical data. In particular it did not exploit the vanishing of the inhomogeneous pressure perturbation as a possible means of escape. 

As we will show below, the class of homogeneous theories that evade the LIGO/Virgo constraints is  also somewhat broader than this one particular example. Indeed, Beyond Horndeski loopholes were prematurely ruled out in \cite{Copeland:2018yuh} using decay constraints that did not directly apply (see footnote \ref{prem}), while DHOST theories were not investigated in any capacity.  All this considered, we  expand our analysis and ask again  whether there are any theories that dynamically evade all of the relevant constraints even in the presence of linear inhomogeneities.  

In the EFT formalism the background evolution is already set by the condition \eqref{eq:conserved_stress_energy} and the perturbations of the scalar field are eaten by the metric. 
Working in the same setup as section \ref{sec:EFT_action}, the only propagating scalar field is the Goldstone boson of the broken time diffs, $\pi$. 
Our approach will be to use the free equation of motion for the $\pi$ perturbation to relax the constraints on the various dark energy theories. 
To do that we need to expand the effective action \eqref{eq:DE_action} at quadratic order, focussing only on operators with at least one $\pi$. Doing that we find (see appendix \ref{app:EFT_action} for further details)
\be\label{eq:quadratic_pi_action}
\begin{split}
S^{(2)}_\pi = \!\! \int \!\! d^4x \, a^3 & \Big\{  \( c + 2 m_2^4 \) \dot\pi^2 + \( 2 c - 3 m_3^3 \dot H \)  \dot \pi \pi - \( m_3^3 + m_4^2 \)\dot \pi {\d^2\pi}/a^2 - 3 m_3^3 \dot \pi \dot \Psi -6 c \dot \pi \Psi \\ 
&- 4 m_4^2 \dot \pi {\d^2 \dot \Psi} / {a^2} - \( 2 c + 4 m_2^4 - 3 H m_3^3 \) \dot \pi \Phi + \Big[3 \dot H \( c + 2 m_2^4 \dot H \) + 3 H \( \dot c - 2 M_*^2  \dot f \dot H \) + \ddot c \\
& - 3/2 M_*^2 \dot f \ddot H \Big] \pi^2 - 4 m_4^2 \dot H \pi \d^2 \pi / a^2 - 3 M_*^2 \dot f \pi \ddot \Psi - 12 \( M_*^2 \dot f H -m_4^2 \dot H \) \pi \dot \Psi - 6 \( 3 H c + \dot c \) \pi \Psi \\
& - 2 M_*^2 \dot f \pi \d^2 \Psi / a^2 - 3 M_*^2 \dot f H \pi \dot \phi + \[ 6 H \( c - 2 M_*^2 \dot f H \) + 3H \( m_3^3 + 4 H m_4^2 -2 M_*^2 \dot f \) \] \pi \Phi \\
& - M_*^2 \dot f \pi \d^2 \Phi /a^2 + 4 m_4^2 \dot \Psi \d^2 \pi /a^2 + \( m_3^3 + 4 H m_4^2 + 4 \widetilde m_4^2 \) \Phi \d^2 \pi /a^2 \Big\} \,. 
\end{split} 
\ee
\subsection{Homogeneous configurations}  \label{sec:new_ths}
We begin by focusing on homogeneous configurations by neglecting any spatial gradients - we will switch them back on in section \ref{sec:perts}. We can easily  obtain the homogeneous free equation of motion for $\pi$, although it is convenient to express this and other dynamical quantities  in terms of  gauge invariants. According to how the homogeneous perturbations $\Phi$, $\Psi$ and $\pi$ transform under a time diff, we can define the following gauge invariant variables,
 \be
 X \equiv \Phi - \dot \pi \,, \qquad Y \equiv H \Phi + \dot H \pi + \dot \Psi \,.
 \ee
 The free equation of motion of $\pi$ becomes
 \bes \label{eq:EoM_pi}
 \Big[ 6 H \( c + 2 m_2^4 \) + 2 \dot c + 3 \dot H &\( m_3^3 - M_*^2 \dot f \) + 4 \dot {(m_4^4)} \Big] X + 2 \( c + 2 m_2^4\) \dot X \\ 
  &+ 3 \[ 2 c + H \( 3m_3^3 - 4 M_*^2 \dot f \) + 4 m_4^2 \dot H + \dot{( m_3^3 )} \] Y + 3 \( m_3^3 - M_*^2 \dot f\) \dot Y = 0
 \end{split}
 \ee
If we solve for $Y$ and plug the solution into the expression of $\alpha_T$, eq.~\eqref{eq:alpha_t}, we get
 \be
 \alpha_T = m_4^2 + \alpha_{\dot Y} \dot Y  + \alpha_{\dot X} \dot X + \alpha_X X \,,
 \ee
 where
 \bea
 \label{eq:alpha_dotY}
\alpha_{\dot Y} &=& \frac{2 m_6 \( m_3^3 - M_*^2 \dot f \)}{2 c + H \( 3 m_3^3 - 4 M_*^2 \dot f \) + 4 m_4^2 \dot H + \dot{(m_3^3)}} \,, \\
 \label{eq:alpha_dotX}
\alpha_{\dot X} &=& - 2 \widetilde m_6 + \frac{4 m_6 \(c + M_*^2 \dot f \)}{3 \[ 2 c + 3 H m_3^3 - 4 H M_*^2 \dot f + 4 \dot H m_4^2 + \dot{(m_3^3)} \] } \,, \\
 \label{eq:alpha_X}
\alpha_X &=& 2 \( m_5^2 - \widetilde m_4^2 - H \widetilde m_6 - \dot{\widetilde m}_6 \) + \frac{2 m_6 \[ 6 H \(c + 2 m_2^4 \) + 2 \dot c + 3 \dot H \( m_3^3 - M_*^2 \dot f\) + 4 \dot {(m_2^4)}\]}{3 \[ 2 c + 3 H m_3^3 - 4 H M_*^2 \dot f + 4 \dot H m_4^2 + \dot{(m_3^3)} \]} \,.
\eea 
This expression for $\alpha_T$ knows about the vanishing of the scalar equation of motion. Therefore, to have $\alpha_T = 0$ in any homogeneous background, the above equations must vanish independently. We can do that either by setting $m_4=0$, $\widetilde m_4^2 = m_5^2$ and  $m_6 = \widetilde m_6 = 0$ (and so we are back to the results of sec.~\ref{sec:Paolos_argument}), or we can set
\be \label{eq:new_ths_1}
m_4 ^2 = 0\,, \qquad m_3^3   = M_*^2 \dot f  \,, \qquad m_6   = \frac 32 \widetilde m_6 \, \frac{ 2 c +M_*^2 \( \ddot f - H \dot f \) }{\( c + 2 m_2^4 \)} \,,
\ee
and require 
\be
\label{eq:new_ths_2}
m_5^2 = \widetilde m_4^2 - 2 H \widetilde m_6 + \dot{\widetilde m}_6 - \widetilde m_6 \d_t \log(c+2m_2^4)
\ee
The above three equations identify a new class of dark energy theories that evade the LIGO/Virgo constraints on homogeneous backgrounds. To ensure that our calculations are correct let us check whether or not  the rescued theory of \cite{Copeland:2018yuh} satisfies the conditions (\ref{eq:new_ths_1}) and (\ref{eq:new_ths_2}).
Using the results of \cite{Gubitosi:2012hu} it is easy to write the rescued theory in the EFT language, 
\bes \label{eq:rescued_th_EFT_params}
& M_*^2 f = 2 \kappa_G - 6 H \mu \,, \qquad 
c = -3 \mu \( H \dot H - \ddot H \) +\frac{\nu e^W \( W' X + 2 \ddot \phi \)}{2 H X \sqrt{-X}}\,, \qquad
m_2^4 = c + \frac 34 M_*^2 \(\ddot f -H \dot f \) \,,\\
& m_3^3 = - 6 \dot H \mu \,, \qquad
m_4^2 = \widetilde m_4^2 = \frac{3\mu}{2\sqrt{-X}} \( 2 \ddot \phi - 2 H \sqrt{-X} + X W'(\phi) \) \,, \qquad
m_5 = 0 \,, \quad 
m_6 = \widetilde m_6 = -3\mu \,.
\end{split}
\ee
First of all notice that the couplings $m_4^2$ and $\widetilde m_4^2$ are not  {identically} zero, but  proportional to the $\phi$ equation of motion (see eq.~\eqref{eq:EoM_rescued}) and so vanishing on shell. We can also check that the parameters of eq. \eqref{eq:rescued_th_EFT_params} satisfy eq. \eqref{eq:new_ths_1}. 
Finally, eq.~\eqref{eq:new_ths_2} is satisfied, again, using the $\phi$ equation of motion \eqref{eq:rescued_th_EFT_params}, completing the check.  It is important to point out that eq \eqref{eq:new_ths_1} and eq \eqref{eq:new_ths_2} yield  a broader class of dark energy theories to the one proposed in \cite{Copeland:2018yuh} . This is because it also retains a rescued class of Beyond Horndeski theories prematurely ruled out in \cite{Copeland:2018yuh}.

\subsection{What about DHOST?}
DHOST theories were not considered in any capacity in \cite{Copeland:2018yuh}. However, the EFT formalism is well adapted to include them.  To this end we  supplement the EFT action \eqref{eq:DE_action} with the following operators, corresponding to DHOST corrections \cite{Langlois:2015cwa, DHOST2, Crisostomi:2019yfo}:
\be 
\label{eq:DHOST_action}
S_{DHOST} = \int d^4x \, \sqrt{-g} \[4\beta_1 \delta K V + \beta_2 V^2 + \beta_3 a_i a^i  \] \,,
\ee 
where $V \equiv (\dot{N} - N^i \partial_i N)/N$ and $a_i \equiv \partial_i N /N$. The action $S \equiv S_{EFT}+S_{DHOST}$, contains all the possible independent operators that start at quadratic level, with at most two derivatives acting on the metric, \cite{Langlois:2015cwa,  Bordin:2017hal}.
Since $S_{DHOST}$ contains operators with one more derivative acting on the fields, in general they propagate more than one scalar degree of freedom. 
Therefore, one has to impose the following degeneracy conditions, that ensure the theory describes only one scalar  degree of freedom \cite{Langlois:2015cwa, DHOST2, Crisostomi:2019yfo},
\be
\beta_2 = -6\beta_1\,, \qquad \beta_3 = -2\beta_1 \[ 2\(1-2 \frac{\widetilde m_4^2}{M_*^2f}\)+\beta_1\]\,.
\ee
Expanding the action \eqref{eq:DHOST_action} in perturbations, we note that these new operators do not lead to changes in the expression for the  speed  of gravitational waves, $c_T$. However, they do contribute to the free scalar equation of motion and can therefore change the expression for $\alpha_T$, indirectly, once we have evaluated it on-shell. 
To investigate this further, let us  expand the action to second order. Since we are interested only in the equation of motion for $\pi$, we consider only those terms which are proportional to $\pi$ or its derivatives. This gives, 
\bes
S_{DHOST} = \int d^4x \, a^3 \Big[ 8 \beta_1  \Big( 3 \dot H \ddot \pi \dot \pi +3 \ddot \pi \dot \Psi &+ 3 H \ddot \pi \Phi + \ddot \pi \d^2 \pi /a^2 - 3 \dot H \dot \pi \dot \Phi - \Phi d^2\pi/a^2 \Big) \\
& + 4 \beta_2 \(\ddot \pi^2 - 2 \ddot\pi \dot \Phi \) + 4 \beta_3  \(\d_i\dot\pi \)^2/a^2 - 2 \d_i\dot\pi \d_i \Phi /a^2 \Big] \,.
\end{split}
\ee
Again, we compute the homogeneous equation of motion of $\pi$ and we express it in terms of the gauge invariant quantities $X$ and $Y$. After solving for $Y$,  and plugging the solution into eq.~\eqref{eq:alpha_t},  we obtain, 
\be
\alpha_T = m_4^2 + \alpha_{\dddot X} \dddot X + \alpha_{\ddot X} \ddot X + \alpha_{\dot X} \dot X + \alpha_X X + \alpha_{\dot Y} \dot  Y \,.
\ee
The functions $\alpha_{\dddot X}\,, \ \alpha_{\ddot X} \,, \  \alpha_{\dot X} \,, \ \alpha_X \,, \  \alpha_{\dot Y}$ are given in app.~\ref{app:DHOST_alpha_t}. Here we just use the fact that
$\alpha_{\dddot X}$ and  $\alpha_{\ddot X}$ both vanish if and only if  $m_6  \beta_1=0$. 
If we want to have a non-trivial DHOST theory with $\beta_1 \not = 0$ we must therefore impose $m_6 = 0$. 
Inspecting equations (\ref{eq:first_app}) to (\ref{eq:last_app}), it is easy to see that this choice implies $\widetilde m_6 = 0$ and $m_5^2 =  \widetilde m_4^2$. In other words, we return to the results of sec.~\ref{sec:Paolos_argument}. We conclude that DHOST theories do not lead to any new results - we need to switch them off in order to escape the conclusions of \cite{Creminelli:2017sry}.

 \subsection{Adding inhomogeneities - an intuitive argument}

We now ask if the class of theories of sec.~\ref{sec:new_ths} continue to satisfy the LIGO/Virgo constraints, even in the presence of  inhomogeneous perturbations.
As a warm up to the main event we provide an intuitive argument to illustrate why this is unlikely. However, we emphasize that this argument will not make use of all of the dynamical information and we stress the importance of a more detailed analysis carried out in the next section.  

To develop our intuition, note that an inhomogeneous perturbation introduces local curvature and so locally,  the universe looks like a curved FRW universe. 
 We can then perform the same analysis of sec.~\ref{sec:new_ths} expanding the action \eqref{eq:DE_action} around the following metric,
\be
ds^2 = - dt^2 + a^2 \( \delta_{ij} + k \frac{x_i \, x_j}{1 - k |\vec x|^2} \) dx^i dx^j \,, 
\ee
with $k \not = 0$. The extrinsic curvature $K_{\mu\nu}$ is not affected by $k$. However,  on the background level, the 3-dimensional Ricci tensor goes as
\be
{}^{(3)} R_{\mu\nu} = \frac {2 k}{a^2}h_{\mu\nu} \,.
\ee
where we recall that $h_{\mu\nu}$ is the induced metric on the spatial slice. Glancing at the form of eq. \eqref{eq:DE_action}, this implies that only the operators proportional to $\widetilde m_4^2$ and $\widetilde m_6$ are affected by the change of the background. 
Let us focus on the latter operator. 
Because of the curved background it now starts at quadratic order in perturbations, 
\be
\widetilde m_6 \delta g^{00} \delta \mathcal G_2 = \widetilde m_6 \delta g^{00} \( \delta K ^\mu _\nu {}^{(3)} R_\mu^\nu - \frac 12 \delta K {}^{(3)} R \) = - \widetilde m_6 \frac{k}{a^2} \delta g^{00} \delta K \,.
\ee
Again, comparing with eq. \eqref{eq:DE_action}, we clearly see that the effect of $\widetilde m_6$, in the presence of curvature, is to shift the coefficient  $m_3^3 \to m_3^3 - 2 k /a^2 \widetilde m_6$.
As a consequence it changes the expression \eqref{eq:alpha_t}  for $\alpha_T$, and, in particular,  the form of  $\alpha_{\dot Y}$. Eq.  \eqref{eq:alpha_dotY}  now becomes
\be
\alpha_{\dot Y} \propto m_3^2 - M_*^2 \dot f - \frac{2k}{a^2} \widetilde m_6 \,.
\ee
This must vanish for any value of $k$. This adds an additional requirement, forcing $\widetilde m_6$ which in turns forces $m_6 = 0$ and $m_5^2 = \widetilde m_4^2$.
Once again we return to the results of \cite{Creminelli:2017sry} and the loophole has been closed, at least intuitively.

\subsection{Adding inhomogeneities - a detailed argument} \label{sec:perts}
We are now ready to perform a more detailed and careful analysis of the LIGO/Virgo constraints in the presence of linearised inhomogeneities. In other words, we restore the spatial gradients.  Crucially, however, there is another consideration that we will take into account that could  help in relaxing the very stringent bounds on the EFT coefficients:  the fact that matter perturbations are pressureless.  This is equivalent to saying that the diagonal part of Einstein's Equations is equal to the stress energy tensor of the dark energy field, or in other words, we demand that the variation of $S_{EFT}$ w.r.t.~$\Psi$ is zero.

The expression we obtain is quite complicated but can be simplified by exploiting the huge separation of scales in the problem at hand.
We are interested in scalar perturbations with the typical size of a galaxy, at the the same time, GWs have a typical length of $\lambda_{GW} \sim 10^3 km$.
Therefore we can exploit the fact that $k_{GW}\sim\lambda_{GW}^{-1} \gg k_s \sim r_{gal}^{-1} \gg H_0$. 

The expression for $\alpha_T$ still takes the form of eq. \eqref{eq:alpha_t}.
As our interest is only  in  those theories that exploited a dynamical loophole at the homogeneous level, we parametrise them completely in terms of $\Lambda$, $c$, $\widetilde m_4$ and $\widetilde m_6$ by using eqs.~\eqref{eq:new_ths_1} and \eqref{eq:new_ths_2}.  We will assume this is the case for the remainder of this section and evaluate $\alpha_T$ accordingly. 

 For these loophole theories,  we now  compute the equation of motion for $\pi$, including spatial gradients \eqref{full_pi_eom}, and the $\Psi$ equation of motion, which is also assumed to vanish in the absence of an inhomogeneous pressure perturbation. We can simplify the latter by neglecting all the terms proportional to $H$ and keeping only those with the lowest number of derivatives, consistent with the hierarchy of scales described above. This gives 
\be\label{eq:pressure-less_eom}
 \Lambda \( \Phi - 3\Psi \) - 2 \dot c \pi + c \( \Phi + 3\Psi -2\dot \pi \) - M_*^2 \[ \dddot f \pi +\ddot f \(\Phi -2 \Psi + \dot \pi \) + \dot f \(\dot \Phi -2 \dot \Psi\)\] = 0\,.
\ee
Assuming $\dot f \neq 0$, we solve the scalar equation of motion for $Y$ and  eq.~\eqref{eq:pressure-less_eom}  for $\dot\Psi$ and substitute back into the appropriate expression for $\alpha_T$. Since we are restricting to loophole theories, we expect this to vanish up to spatial gradients, which is indeed the case. Specifically, we get, 
\be
\alpha_T =\frac{1}{a^2} \[ (\alpha_1 H^{-1}) \  \d^2 \pi + ({\alpha_2}{H^{-2}}) \ \d^2\dot\pi + ({\alpha_3}{H^{-2}}) \ \d^2 \Phi + ({\alpha_4}{H^{-3}}) \, \d^2\dot\Phi + ({\alpha_5}{H^{-2}}) \ \d^2 \Psi\],
\ee
where
\begin{eqnarray}
\label{A_coeff} \alpha_1 \!\! &=& \!\! -\frac{2 H \widetilde m_6} {M_*^4 f \dot f (c+2 m_2^4) } \left\{\widetilde m_4^2 \[2 \dot c+M_*^2 \(\dddot f-2 \dot f (\dot H+H^2)\)\]+M_*^2 \dot f \[2 c+M_*^2 \ddot f-2 H \(M_*^2 \dot f+ (\widetilde m_4^2)\dot{} \)\]\right\}, \nonumber \\ \\
\label{B_coeff} \alpha_2  \!\! &=& \!\! -\frac{2 H^2 \ \widetilde m_4^2 \ \widetilde m_6}{M_*^4 f \dot f  (c+2 m_2^4) } \left(2 c+M_*^2 \ddot f\right)\,, \\ \nonumber \\
\label{C_coeff} \alpha_3  \!\! &=& \!\! \frac{2 H^2 \ \widetilde m_4^2 \ \widetilde m_6}{M_*^4 f \dot f   (c+2 m_2^4)} \[ c+M_*^2 \(-\ddot f+2 H \dot f+3 f H^2 \)-\rho_m \]\,, \\ \nonumber \\
\label{D_coeff} \alpha_4  \!\! &=& \!\! -\frac{2 H^3 \ \widetilde m_4^2 \ \widetilde m_6}{M_*^2 f (c+2 m_2^4)}\,, \\ \nonumber \\
\label{E_coeff} \alpha_5  \!\! &=& \!\! \frac{2 H^2 \widetilde m_6}{M_*^4 f \dot f (c+2 m_2^4) }  \[ \widetilde m_4^2 \( 3 c +2 M_*^2 \ddot f-2 M_*^2 H \dot f  - 9 M_*^2 f H^2 + 3 \rho_m \) + M^4 \dot f^2 - 2 M_*^2 \,\dot f \, (\widetilde m_4^2)\dot{} \]  \,.
\end{eqnarray}
In deriving the above expressions we have also  used the Friedmann equation \eqref{eq:fried_eq} to express $\Lambda$ in terms of $c$ and $\rho_m$, the matter density.
We clearly see that to robustly set $\alpha_T$ to zero we must demand that either  $\widetilde m_4$ or $\widetilde m_6$ are vanishing.
However, from eq.~\eqref{E_coeff}  $\widetilde m_4=0$ implies $\dot f =0$  which contradicts our earlier assumption, so we must have $\widetilde m_6 =0$. This now implies, via eqs.~\eqref{eq:new_ths_1}-\eqref{eq:new_ths_2}, that $m_6=0$ and $\widetilde m_4 = m_5$,  thus going back to the results of \cite{Creminelli:2017sry} and eliminating the loophole. 

The only possible way out is if $f=const$ (we set it to unity for simplicity). In this situation the square bracket in the $\Psi$ equation of motion eq.~\eqref{eq:pressure-less_eom} is identically zero. Instead of solving for $\dot \Psi$ as we did previously, we now solve for $\Psi$ and plug that into $\alpha_T$, arriving at the following expression,
\be
\alpha_T = \frac{1}{a^{2}} \[({\widebar \alpha_1}{H^{-1}}) \ \d^2 \pi + ({\widebar \alpha_2}{H^{-2}}) \ \d^2\dot \pi + ({\widebar \alpha_3}{H^{-2}})\ \d^2 \Phi + ({\widebar \alpha_4}{H^{-3}}) \  \d^2 \dot \Psi\],
\ee
where
\bea
\widebar \alpha_1 \!\! &=& \!\! \frac{4 H \widetilde m_6}{3 M_*^2  (c+2 m_2^4) \left(2 c-3 M_*^2 H^2+\rho_m\right)} \Big\{ \widetilde m_4^2 \(-2 H \dot c+3 \rho_m \dot H+3 H^2 \rho_m\)+(\widetilde m_4^2)\dot{} \(3 H \rho_m-2 \dot c\) \nonumber \\
&& +c \[9 M_*^2 H^2+6 \widetilde m_4^2 (\dot H+H^2 )+6 H (\widetilde m_4^2)\dot{}-3 \rho_m\] -6 c^2-9 M_*^2 H^2 \[ \widetilde m_4^2 (\dot H+H^2)+H (\widetilde m_4^2)\dot{} \] \Big\}, \nonumber \\ \\
\widebar \alpha_2 \!\! &=& \!\! -8 H^2 \ c\  \widetilde m_6 \ \frac{H \widetilde m_4^2+(\widetilde m_4^2)\dot{}}{3 M_*^2 (c+2 m_2^4) (2 c-3 M_*^2 H^2+\rho_m )}\,, \\ \nonumber \\
\widebar \alpha_3 \!\! &=& \!\! 4 H^2 \widetilde m_6 \ \frac{6 c H \widetilde m_4^2-\left(3 M_*^2 H^2-\rho_m\right) \[2 H
   \widetilde m_4^2-(\widetilde m_4^2)\dot{}\]}{3 M_*^2 (c+2 m_2^4) (2 c-3 M_*^2 H ^2+\rho_m )}\,, \\ \nonumber \\
\widebar \alpha_4 \!\! &=& \!\! -\frac{4 H^3 \ \widetilde m_4^2 \ \widetilde m_6}{ M_*^2 (c+2 m_2^4)}\,. \\
\eea
We can achieve $\alpha_T = 0$ in one of two ways.  The first is to set $\widetilde m_6 =0 $, leading us back to the results of \cite{Creminelli:2017sry} and closing the loophole.  The other possibility is to set $\widetilde m_4=0$ and $c=0$. However, as stated in the text after  eq. \ref{eq:conserved_stress_energy}, this corresponds to dark energy driven by a cosmological constant, as in the standard cosmological model, and not some large distance modification of gravity.  Although this set up would differ from the standard $\Lambda$CDM scenario at the level of perturbations, we don't think the scalar has any right to be called a genuine dark energy field. 

\section{Conclusions} \label{sec:conc}
After an exhaustive treatment taking into account all of the relevant dynamical information the conclusion is clear: there are no dark energy loopholes that survive the necessary constraints up to and including leading order cosmological perturbations.  So, a decade and a half since one of us reviewed the state of play \cite{edrev}, what now for dark energy?

Within the playground of scalar tensor theories, there is very little room for manoeuvre. Concerns about the low cut off notwithstanding \cite{claudia}, there are very few models of genuine  cosmological interest that survive the purge carried out by LIGO and Virgo. One class of models that still survive are the so-called KGB set-ups \cite{KGB}.  Although there are some special cases, such as the cubic galileon, that are ruled out by other cosmological constraints \cite{iswgal}, there is sufficient freedom in the KGB potentials to still extract genuine dark energy candidates \cite{kgbgood, kgbgood2}. Beyond that there is less room for optimism, Take, for example, the chameleon models \cite{cham1, cham2}. Although these are compatible with the gravitational wave data, the chameleon by itself cannot be considered a bonafide dark energy field as other constraints on its mass render it irrelevant on Hubble scales  \cite{chamno}. In such scenarios, dark energy must be driven by a cosmological constant and the case for modified gravity with chameleons is less strong.   Similarly, we could  consider generic Horndeski theories where the higher order operators are suppressed by scales larger than Hubble, and we would have no concerns with the speed of propagation of gravitational waves. However, once again, in such a set up, dark energy must be driven by a cosmological constant and the case of modified gravity is weakened.  

Perhaps this is pointing us towards a much less flamboyant origin of dark energy, corresponding to a cosmological constant, or a weakly coupled quintessence field in slow roll.  However, even these scenarios are now being challenged by the so-called swampland criteria \cite{swamp1, swamp2}.  There should be some caution here.  Contrary to the claim made in \cite {swampcos} a slowly rolling quintessence field will  not fall victim to the distance conjecture after an order one excursion in units of the Planck mass, but after  $\mathcal{O}(100)$.  This is because the heavy moduli initially have their masses set by some high ultra-violet scale, $M_{UV}$ which is then modified as  $M_{UV}e^{-\mathcal{O}(1)\Delta \phi/M_p}$  as the dark energy field rolls a distance $\Delta \phi$. To bring their scale down to Hubble today, and contaminate quintessence, we need a very large exponential suppression and a large number of Planck excursions. In our opinion, modelling dark energy within string theory is now more important than ever, especially in light of the constraints coming from LIGO and Virgo. 
For example, some attempts to incorporate dark energy in supergravity can be found in  \cite{Brax:1999gp,Brax:1999yv,Copeland:2000vh} and more recently in  \cite{cesc, stuck}.

\acknowledgments

We would like to thanks Paolo Creminelli, Giovani Tambalo, Vicharit Yingcharoenrat and Alex Vikman for useful discussions. AP was funded by a Leverhulme Research Project Grant. LB, EC and AP were funded by an STFC Consolidated Grant Number ST/P000703/1.

\appendix
\section{Expansion of the effective action}
\label{app:EFT_action}

As our aim is to obtain the equation of motion of $\pi$ and derive the action to first order in scalar and second order in tensor perturbations, we ignore all the terms quadratic in $\Phi$ and $\Psi$. 
To restore the $\pi$ field one needs to perform the time diff
\be
t \to t + \pi(t, \x) \,.
\ee
Under this transformation, any function of time $f$ changes up to second order as
\be
f \to f + \dot f \pi + \frac{\ddot f}{2} \pi^2 + \mathcal O (3)\,,
\ee
while tensors transform as
\be\label{app_eq:metric_transf}
T^{\mu\nu} \to (\delta^\mu_\alpha + \delta^\mu_0 \pi ) (\delta^\nu_\beta + \delta^\nu_0 \pi) \ T^{\alpha \beta}\,,
\ee
from which we can derive the transformations of the different ADM components of the metric. The metric $g^{\mu\nu}$ is decomposed as
\be
g^{00} = - \frac 1 {(1+\delta N)^2}\,, \qquad g^{0i} = \frac {N^i}{(1+\delta N)^2}\,, \qquad g^{ij} = h^{ij} - \frac{N^i N^j}{(1 + \delta N)^2}\,.
\ee
From eq.~\eqref{app_eq:metric_transf} we get
\bea
&& \delta N \to \delta N - \dot\pi + \dot \pi^2 - \dot \pi \delta N + N^i \d_i \pi + \frac 1 {2 a^{2}}\, \d_i \pi \d_j \pi + \mathcal O (3) \,, \\
&& N^i \to N^i(1 - \dot \pi) + (1 + \delta N)^2 h^{ik} \d_k \pi + \mathcal O (3) \,, \\
&& h_{ij} \to h_{ij} - N_i \d_j \pi - N_j \d_i \pi - \d_i \pi \d_j \pi + \mathcal O (3) \,.
\eea
Making use of these relations, together with the derivative transformations,
\be
\d_0 \to (1 - \dot \pi + \dot \pi^2) \d_0 + \mathcal O (3) \,, \qquad \d_i \to \d_i  - (1 - \dot \pi) \d_i \pi \d_0 + \mathcal O (3) \,,
\ee
we can derive the transformations of the extrinsic curvature tensor and 3-dimensional Ricci scalar 
\bea 
&& \delta K_ i ^{\,j} \to \delta K_ i ^{\,j} - \dot H \pi \delta_i ^{\,j} - h^{ik} \d_k \d_j \pi + \mathcal O (2) \,, \\
&& {}^{(3)} R \to {}^{(3)} R - 2 \dot h^{ij} \d_i \d_j \pi + \mathcal O (2) \, \\
\eea
To compute the action we also need to expand the operators before performing the time diff.
We work in Newtonian gauge, so we have
\be
\delta N = \Phi \,, \qquad N^i = 0 \,, \qquad h_{ij} = a^2 \[ \delta_{ij}(1 - 2 \Psi) + \gamma_{ij} \] + \mathcal O (2) \,.
\ee
and so
\bea
&& \delta K_i^{\ j} = - \( \dot \Psi + H \Phi \) \delta _{i}^{\ j} + \frac 12 (1 - \Phi) \delta ^{jk}\dot \gamma_{ik} + \mathcal O (2)\,, \\
&& {}^{(3)} R = -4 a^{-2} \d^2 \Psi + \mathcal O (2) \,, \\
&& R = 6 \Big(2 H^2 + \dot H \Big) - 2 \[ 6 \Big(2 H^2 + \dot H \Big) \Phi + 3 H \dot \Phi + 12 H \dot \Psi + 3 \ddot \Psi + \frac{\d^2}{a^2} \Big(\Phi - 2 \Psi \Big)\] + \mathcal O (2) \,. 
\eea
 It is now straightforward to expand the effective  action up to second order in gravitons and first order in scalar  perturbations,  
\begin{eqnarray} \label{eq:full_eft_action}
S[\gamma] &=& \int d^4x \, a^3 \Big\{ \frac{M_*^2 f}{8}  \Big[ \(1-\Phi - 3 \Psi + \dot f / f \) \dot \gamma_{ij}^2 - a^{-2} \(1-\Phi+\Psi + \dot f / f \) (\d_k\gamma_{ij})^2 \nonumber \\
&& + 4a^{-2} \d_j \Big( \d_i \(2\Psi-\Phi\) \gamma_{ik} \Big) \gamma_{kj} \Big] + \frac{m_4^2}4 \Big(\dot \gamma_{ij}^2 - 4a^{-2}\d_i\d_j\pi\gamma_{ik}\gamma_{kj} \Big) + \frac{a^{-2}}{4} \widetilde m_4^2 \(\Phi - \dot \pi\) \(\d_k \gamma_{ij}\)^2 \nonumber \\
&&  + \frac{m_5^2}{4} \(\Phi - \dot \pi \) \dot \gamma_{ij}^2 - \frac{m_6}{4} \[ \(H \Phi+\dot H \pi +\dot\Psi + \frac{\d^2\pi}{a^2} \) \dot\gamma_{ij}^2 - \frac 32 a^{-3} \d_i\d_j \pi \dot \gamma_{ik} \dot \gamma_{jk} \] \nonumber \\
&& + \frac{a^{-2}}{4} \widetilde m_6 \Big[ 2 \d_k \(\Phi - \pi\) \dot\gamma_{ij}\d_k\gamma_{ij} - \Big( H\(\Phi-\dot\pi\) + (\dot\Phi - \ddot \pi)\Big) \(\d_k\gamma_{ij}\)^2 \Big] - \frac{a^{-2}}{4} \dot{\widetilde m}_6 \(\Phi - \dot\pi \) \(\d_k \gamma_{ij}\)^2 \Big\} \nonumber \\
\end{eqnarray}
To recover the limit of General Relativity, we set  all the $m_i$ couplings to zero and choose $f=1$.

\section{Expression for $\alpha_T$ in DHOST}
\label{app:DHOST_alpha_t}

Including also the DHOST operators, the coefficient $\alpha_T$ is given by
\be
\alpha_T = m_4^2 + \alpha_{\dddot X} \dddot X + \alpha_{\ddot X} \ddot X + \alpha_{\dot X} \dot X + \alpha_X X + \alpha_{\dot Y} \dot  Y \,,
\ee
where
{\footnotesize
\be \label{eq:first_app}
\alpha_{\dddot X} = \frac{16 f \ m_6 \ \beta_1^2}
{ \[c + 3Hm_3^3 + (m_3^3) \dot {} \, \]/M_*^2+ 2 H \[(3 \beta_1-2) \dot f+3 f \dot \beta_1 \]+ \[ \beta_1 \ddot f +2 \dot f \dot \beta_1 + f \Big(3 \beta_1 \dot H + \ddot \beta_1\Big)\]+9 f H^2 \beta_1 } 
\ee
\be
\alpha_{\ddot X} = \frac{32 m_6 \beta_1 \[ \beta_1 \Big( \dot f + 3 f H \Big) + 2 f \dot \beta_1\] }{ \[c + 3Hm_3^3 + (m_3^3) \dot {} \, \]/M_*^2+ 2 H \[(3 \beta_1-2) \dot f+3 f \dot \beta_1 \]+ \[ \beta_1 \ddot f +2 \dot f \dot \beta_1 + f \Big(3 \beta_1 \dot H + \ddot \beta_1\Big)\]+9 f H^2 \beta_1 }
\ee
\be
\begin{split} 
\alpha_{\dot X} = & - \frac 23 \frac{ \widetilde m_6 - m_6 \left\{[2 c + 4 m_2^4] / M_*^2 +  24 \beta_1 \[ \beta_1\ddot f + 2 \dot f  \(3 H \beta_1+ 2 \dot \beta_1 \)\] \right\}}{ \[c + 3Hm_3^3 + (m_3^3) \dot {} \, \]/M_*^2+ 2 H \[(3 \beta_1-2) \dot f+3 f \dot \beta_1 \]+ \[ \beta_1 \ddot f +2 \dot f \dot \beta_1 + f \Big(3 \beta_1 \dot H + \ddot \beta_1\Big)\]+9 f H^2 \beta_1 } \\
&+ \frac 23 \frac{3 f \ m_6 \[ (24 \beta_1 - 1) \beta_1 \dot H + 16 \beta_1 \ddot \beta_1 + 96 H \beta_1 \dot \beta_1 + 72 H^2 \beta_1^2 + 16 \dot \beta_1^2 \] }{\[c + 3Hm_3^3 + (m_3^3) \dot {} \, \]/M_*^2+ 2 H \[(3 \beta_1-2) \dot f+3 f \dot \beta_1 \]+ \[ \beta_1 \ddot f +2 \dot f \dot \beta_1 + f \Big(3 \beta_1 \dot H + \ddot \beta_1\Big)\]+9 f H^2 \beta_1}
\end{split}
\ee
\be
\alpha_X = \frac{2 m_6 \[ 2 \dot c + 6 H ( c + 2 m_2^4 ) + 3 \dot H \( m_3^3 - M_*^2 \dot f \) + 4 (m_2^4) \dot {} \] }{3 \left\{ \[c + 3Hm_3^3 + (m_3^3) \dot {} \, \]/M_*^2+ 2 H \[(3 \beta_1-2) \dot f+3 f \dot \beta_1 \]+ \[ \beta_1 \ddot f +2 \dot f \dot \beta_1 + f \Big(3 \beta_1 \dot H + \ddot \beta_1\Big)\]+9 f H^2 \beta_1 \right\}}
\ee
\be
\alpha_{\ddot Y} = \frac{2 f m_6 \beta_1}{\[c + 3Hm_3^3 + (m_3^3) \dot {} \, \]/M_*^2+ 2 H \[(3 \beta_1-2) \dot f+3 f \dot \beta_1 \]+ \[ \beta_1 \ddot f +2 \dot f \dot \beta_1 + f \Big(3 \beta_1 \dot H + \ddot \beta_1\Big)\]+9 f H^2 \beta_1}
\ee
\be \label{eq:last_app}
\alpha_{\dot Y} = \frac{2 m_6 \left[ (2 \beta_1 -1) \dot f + 2 f  \left(3 H  \beta_1 + \dot \beta_1 \right)+ m_3^3/M_*^2 \right]}{\[c + 3Hm_3^3 + (m_3^3) \dot {} \, \]/M_*^2+ 2 H \[(3 \beta_1-2) \dot f+3 f \dot \beta_1 \]+ \[ \beta_1 \ddot f +2 \dot f \dot \beta_1 + f \Big(3 \beta_1 \dot H + \ddot \beta_1\Big)\]+9 f H^2 \beta_1}
\ee}

\section{Equations of motion for the $\pi$ field including spatial gradients}
From the quadratic action \eqref{eq:quadratic_pi_action} we can derive the following equation of motion for $\pi$, including gradients.
\bea \label{full_pi_eom}
&& -2  \ddot \pi \ (c+2 m_2^4) + 3 \ddot \Psi \ \(m_3^3-M_*^2 \dot f \) -2 \dot \pi \ \left[ \dot c+3 H (c+2 m_2^4)+2 (m_2^4)\dot{} \] + 3 \dot \Psi \ \Big[2 c+H (3 m_3^3-4 M_*^2 \dot f ) \nonumber \\
&& +4 m_4^2 \dot H+(m_3^3)\dot{} \Big] + \dot \Phi \[2 c+3 H ( m_3^3-M_*^2 \dot f )+4 m_2^4 \] + \pi \Big[3 \dot H \( 2 c+H (3 m_3^3-4 M_*^2 \dot f)+4 m_4^2 \dot H+(m_3^3)\dot{} \) \nonumber \\
&& +3 \ddot H ( m_3^3-M_*^2\dot f ) \Big] + \Phi \ \Big[2 \dot c+12 c H+6 \dot H (m_3^3-M_*^2 \dot f )+3 H^2 (3 m_3^3-4 M_*^2 \dot f )+3 H \Big(4 m_4^2 \dot H+4m_2^4 \nonumber \\
&& +(m_3^3)\dot{} \Big)+4 (m_2^4)\dot{} \Big]  + a^{-2} \Big\{ \d^2 \pi \[ 2 c+4 \dot H (\widetilde m_4^2-2 m_4^2)+H \left(m_3^3+4 (\widetilde m_4^2)\dot{}\right)+4 H^2 \widetilde m_4^2+(m_3^3)\dot{} \] \nonumber \\
&&  + 2 \d^2 \Psi \[ M_*^2 \dot f-2 H \widetilde m_4^2-2 (\widetilde m_4^2)\dot{} \] + \d^2 \Phi \[ - M_*^2 \dot f +4 H (m_4^2+\widetilde m_4^2)+m_3^3 \] + 4  \d^2 \dot \Psi \ (m_4^2-\widetilde m_4^2) \Big\} =0 \,. \nonumber \\
\eea


\bibliographystyle{utphys}
\bibliography{my_biblio.bib}

\providecommand{\href}[2]{#2}\begingroup\raggedright\begin{thebibliography}{10}

\bibitem{GW170817}
{\bf LIGO Scientific, Virgo} Collaboration, B.~Abbott {\em et.~al.},
  ``{GW170817: Observation of Gravitational Waves from a Binary Neutron Star
  Inspiral},'' {\em Phys. Rev. Lett.} {\bf 119} (2017), no.~16 161101,
  \href{http://xxx.lanl.gov/abs/1710.05832}{{\tt 1710.05832}}.

\bibitem{GRB170817A}
A.~Goldstein {\em et.~al.}, ``{An Ordinary Short Gamma-Ray Burst with
  Extraordinary Implications: Fermi-GBM Detection of GRB 170817A},'' {\em
  Astrophys. J. Lett.} {\bf 848} (2017), no.~2 L14,
  \href{http://xxx.lanl.gov/abs/1710.05446}{{\tt 1710.05446}}.

\bibitem{Integral}
V.~Savchenko {\em et.~al.}, ``{INTEGRAL Detection of the First Prompt Gamma-Ray
  Signal Coincident with the Gravitational-wave Event GW170817},'' {\em
  Astrophys. J. Lett.} {\bf 848} (2017), no.~2 L15,
  \href{http://xxx.lanl.gov/abs/1710.05449}{{\tt 1710.05449}}.

\bibitem{combo}
{\bf LIGO Scientific, Virgo, Fermi-GBM, INTEGRAL} Collaboration, B.~Abbott {\em
  et.~al.}, ``{Gravitational Waves and Gamma-rays from a Binary Neutron Star
  Merger: GW170817 and GRB 170817A},'' {\em Astrophys. J. Lett.} {\bf 848}
  (2017), no.~2 L13, \href{http://xxx.lanl.gov/abs/1710.05834}{{\tt
  1710.05834}}.

\bibitem{multi-mess}
{\bf LIGO Scientific, Virgo, Fermi GBM, INTEGRAL, IceCube, AstroSat Cadmium
  Zinc Telluride Imager Team, IPN, Insight-Hxmt, ANTARES, Swift, AGILE Team,
  1M2H Team, Dark Energy Camera GW-EM, DES, DLT40, GRAWITA, Fermi-LAT, ATCA,
  ASKAP, Las Cumbres Observatory Group, OzGrav, DWF (Deeper Wider Faster
  Program), AST3, CAASTRO, VINROUGE, MASTER, J-GEM, GROWTH, JAGWAR,
  CaltechNRAO, TTU-NRAO, NuSTAR, Pan-STARRS, MAXI Team, TZAC Consortium, KU,
  Nordic Optical Telescope, ePESSTO, GROND, Texas Tech University, SALT Group,
  TOROS, BOOTES, MWA, CALET, IKI-GW Follow-up, H.E.S.S., LOFAR, LWA, HAWC,
  Pierre Auger, ALMA, Euro VLBI Team, Pi of Sky, Chandra Team at McGill
  University, DFN, ATLAS Telescopes, High Time Resolution Universe Survey,
  RIMAS, RATIR, SKA South Africa/MeerKAT} Collaboration, B.~Abbott {\em
  et.~al.}, ``{Multi-messenger Observations of a Binary Neutron Star Merger},''
  {\em Astrophys. J. Lett.} {\bf 848} (2017), no.~2 L12,
  \href{http://xxx.lanl.gov/abs/1710.05833}{{\tt 1710.05833}}.

\bibitem{Creminelli:2017sry}
P.~Creminelli and F.~Vernizzi, ``{Dark Energy after GW170817 and GRB170817A},''
  {\em Phys. Rev. Lett.} {\bf 119} (2017), no.~25 251302,
  \href{http://xxx.lanl.gov/abs/1710.05877}{{\tt 1710.05877}}.

\bibitem{Sakstein}
J.~Sakstein and B.~Jain, ``{Implications of the Neutron Star Merger GW170817
  for Cosmological Scalar-Tensor Theories},'' {\em Phys. Rev. Lett.} {\bf 119}
  (2017), no.~25 251303, \href{http://xxx.lanl.gov/abs/1710.05893}{{\tt
  1710.05893}}.

\bibitem{Ezquiaga:2017ekz}
J.~M. Ezquiaga and M.~Zumalac\'arregui, ``{Dark Energy After GW170817: Dead
  Ends and the Road Ahead},'' {\em Phys. Rev. Lett.} {\bf 119} (2017), no.~25
  251304, \href{http://xxx.lanl.gov/abs/1710.05901}{{\tt 1710.05901}}.

\bibitem{Baker}
T.~Baker, E.~Bellini, P.~Ferreira, M.~Lagos, J.~Noller, and I.~Sawicki,
  ``{Strong constraints on cosmological gravity from GW170817 and GRB
  170817A},'' {\em Phys. Rev. Lett.} {\bf 119} (2017), no.~25 251301,
  \href{http://xxx.lanl.gov/abs/1710.06394}{{\tt 1710.06394}}.

\bibitem{Amendola}
L.~Amendola, M.~Kunz, I.~D. Saltas, and I.~Sawicki, ``{Fate of Large-Scale
  Structure in Modified Gravity After GW170817 and GRB170817A},'' {\em Phys.
  Rev. Lett.} {\bf 120} (2018), no.~13 131101,
  \href{http://xxx.lanl.gov/abs/1711.04825}{{\tt 1711.04825}}.

\bibitem{Langlois}
D.~Langlois, R.~Saito, D.~Yamauchi, and K.~Noui, ``{Scalar-tensor theories and
  modified gravity in the wake of GW170817},'' {\em Phys. Rev. D} {\bf 97}
  (2018), no.~6 061501, \href{http://xxx.lanl.gov/abs/1711.07403}{{\tt
  1711.07403}}.

\bibitem{Crisostomi}
M.~Crisostomi and K.~Koyama, ``{Vainshtein mechanism after GW170817},'' {\em
  Phys. Rev. D} {\bf 97} (2018), no.~2 021301,
  \href{http://xxx.lanl.gov/abs/1711.06661}{{\tt 1711.06661}}.

\bibitem{Babichev:2017lmw}
E.~Babichev, C.~Charmousis, G.~Esposito-Farèse, and A.~Lehébel, ``{Stability
  of Black Holes and the Speed of Gravitational Waves within Self-Tuning
  Cosmological Models},'' {\em Phys. Rev. Lett.} {\bf 120} (2018), no.~24
  241101, \href{http://xxx.lanl.gov/abs/1712.04398}{{\tt 1712.04398}}.

\bibitem{Kase}
R.~Kase and S.~Tsujikawa, ``{Dark energy in Horndeski theories after GW170817:
  A review},'' {\em Int. J. Mod. Phys. D} {\bf 28} (2019), no.~05 1942005,
  \href{http://xxx.lanl.gov/abs/1809.08735}{{\tt 1809.08735}}.

\bibitem{Copeland:2018yuh}
E.~J. Copeland, M.~Kopp, A.~Padilla, P.~M. Saffin, and C.~Skordis, ``{Dark
  energy after GW170817 revisited},'' {\em Phys. Rev. Lett.} {\bf 122} (2019),
  no.~6 061301, \href{http://xxx.lanl.gov/abs/1810.08239}{{\tt 1810.08239}}.

\bibitem{tonyrev}
T.~Clifton, P.~G. Ferreira, A.~Padilla, and C.~Skordis, ``{Modified Gravity and
  Cosmology},'' {\em Phys. Rept.} {\bf 513} (2012) 1--189,
  \href{http://xxx.lanl.gov/abs/1106.2476}{{\tt 1106.2476}}.

\bibitem{joycerev}
A.~Joyce, B.~Jain, J.~Khoury, and M.~Trodden, ``{Beyond the Cosmological
  Standard Model},'' {\em Phys. Rept.} {\bf 568} (2015) 1--98,
  \href{http://xxx.lanl.gov/abs/1407.0059}{{\tt 1407.0059}}.

\bibitem{Horndeski}
G.~W. Horndeski, ``{Second-order scalar-tensor field equations in a
  four-dimensional space},'' {\em Int. J. Theor. Phys.} {\bf 10} (1974)
  363--384.

\bibitem{Horn2}
C.~Deffayet, X.~Gao, D.~Steer, and G.~Zahariade, ``{From k-essence to
  generalised Galileons},'' {\em Phys. Rev. D} {\bf 84} (2011) 064039,
  \href{http://xxx.lanl.gov/abs/1103.3260}{{\tt 1103.3260}}.

\bibitem{Bhorn}
J.~Gleyzes, D.~Langlois, F.~Piazza, and F.~Vernizzi, ``{Healthy theories beyond
  Horndeski},'' {\em Phys. Rev. Lett.} {\bf 114} (2015), no.~21 211101,
  \href{http://xxx.lanl.gov/abs/1404.6495}{{\tt 1404.6495}}.

\bibitem{Bhorn2}
J.~Gleyzes, D.~Langlois, F.~Piazza, and F.~Vernizzi, ``{Exploring gravitational
  theories beyond Horndeski},'' {\em JCAP} {\bf 02} (2015) 018,
  \href{http://xxx.lanl.gov/abs/1408.1952}{{\tt 1408.1952}}.

\bibitem{Jimenez}
J.~Beltran~Jimenez, F.~Piazza, and H.~Velten, ``{Evading the Vainshtein
  Mechanism with Anomalous Gravitational Wave Speed: Constraints on Modified
  Gravity from Binary Pulsars},'' {\em Phys. Rev. Lett.} {\bf 116} (2016),
  no.~6 061101, \href{http://xxx.lanl.gov/abs/1507.05047}{{\tt 1507.05047}}.

\bibitem{Lombriser1}
L.~Lombriser and A.~Taylor, ``{Breaking a Dark Degeneracy with Gravitational
  Waves},'' {\em JCAP} {\bf 03} (2016) 031,
  \href{http://xxx.lanl.gov/abs/1509.08458}{{\tt 1509.08458}}.

\bibitem{Lombriser2}
L.~Lombriser and N.~A. Lima, ``{Challenges to Self-Acceleration in Modified
  Gravity from Gravitational Waves and Large-Scale Structure},'' {\em Phys.
  Lett. B} {\bf 765} (2017) 382--385,
  \href{http://xxx.lanl.gov/abs/1602.07670}{{\tt 1602.07670}}.

\bibitem{KGB}
C.~Deffayet, O.~Pujolas, I.~Sawicki, and A.~Vikman, ``{Imperfect Dark Energy
  from Kinetic Gravity Braiding},'' {\em JCAP} {\bf 10} (2010) 026,
  \href{http://xxx.lanl.gov/abs/1008.0048}{{\tt 1008.0048}}.

\bibitem{claudia}
C.~de~Rham and S.~Melville, ``{Gravitational Rainbows: LIGO and Dark Energy at
  its Cutoff},'' {\em Phys. Rev. Lett.} {\bf 121} (2018), no.~22 221101,
  \href{http://xxx.lanl.gov/abs/1806.09417}{{\tt 1806.09417}}.

\bibitem{Cheung:2007st}
C.~Cheung, P.~Creminelli, A.~L. Fitzpatrick, J.~Kaplan, and L.~Senatore, ``{The
  Effective Field Theory of Inflation},'' {\em JHEP} {\bf 0803} (2008) 014,
  \href{http://xxx.lanl.gov/abs/0709.0293}{{\tt 0709.0293}}.

\bibitem{Gubitosi:2012hu}
G.~Gubitosi, F.~Piazza, and F.~Vernizzi, ``{The Effective Field Theory of Dark
  Energy},'' {\em JCAP} {\bf 1302} (2013) 032,
  \href{http://xxx.lanl.gov/abs/1210.0201}{{\tt 1210.0201}}.
  [JCAP1302,032(2013)].

\bibitem{Creminelli:2018xsv}
P.~Creminelli, M.~Lewandowski, G.~Tambalo, and F.~Vernizzi, ``{Gravitational
  Wave Decay into Dark Energy},'' {\em JCAP} {\bf 1812} (2018), no.~12 025,
  \href{http://xxx.lanl.gov/abs/1809.03484}{{\tt 1809.03484}}.

\bibitem{Langlois:2015cwa}
D.~Langlois and K.~Noui, ``{Degenerate higher derivative theories beyond
  Horndeski: evading the Ostrogradski instability},'' {\em JCAP} {\bf 1602}
  (2016), no.~02 034, \href{http://xxx.lanl.gov/abs/1510.06930}{{\tt
  1510.06930}}.

\bibitem{DHOST2}
J.~Ben~Achour, D.~Langlois, and K.~Noui, ``{Degenerate higher order
  scalar-tensor theories beyond Horndeski and disformal transformations},''
  {\em Phys. Rev. D} {\bf 93} (2016), no.~12 124005,
  \href{http://xxx.lanl.gov/abs/1602.08398}{{\tt 1602.08398}}.

\bibitem{Crisostomi:2019yfo}
M.~Crisostomi, M.~Lewandowski, and F.~Vernizzi, ``{Vainshtein regime in
  scalar-tensor gravity: Constraints on degenerate higher-order scalar-tensor
  theories},'' {\em Phys. Rev.} {\bf D100} (2019), no.~2 024025,
  \href{http://xxx.lanl.gov/abs/1903.11591}{{\tt 1903.11591}}.

\bibitem{kob}
T.~Kobayashi, M.~Yamaguchi, and J.~Yokoyama, ``{Generalized G-inflation:
  Inflation with the most general second-order field equations},'' {\em Prog.
  Theor. Phys.} {\bf 126} (2011) 511--529,
  \href{http://xxx.lanl.gov/abs/1105.5723}{{\tt 1105.5723}}.

\bibitem{Bordin:2017hal}
L.~Bordin, G.~Cabass, P.~Creminelli, and F.~Vernizzi, ``{Simplifying the EFT of
  Inflation: generalized disformal transformations and redundant couplings},''
  {\em JCAP} {\bf 1709} (2017), no.~09 043,
  \href{http://xxx.lanl.gov/abs/1706.03758}{{\tt 1706.03758}}.

\bibitem{edrev}
E.~J. Copeland, M.~Sami, and S.~Tsujikawa, ``{Dynamics of dark energy},'' {\em
  Int. J. Mod. Phys. D} {\bf 15} (2006) 1753--1936,
  \href{http://xxx.lanl.gov/abs/hep-th/0603057}{{\tt hep-th/0603057}}.

\bibitem{iswgal}
J.~Renk, M.~Zumalac\'arregui, F.~Montanari, and A.~Barreira, ``{Galileon
  gravity in light of ISW, CMB, BAO and H$_0$ data},'' {\em JCAP} {\bf 10}
  (2017) 020, \href{http://xxx.lanl.gov/abs/1707.02263}{{\tt 1707.02263}}.

\bibitem{kgbgood}
S.~Peirone, G.~Benevento, N.~Frusciante, and S.~Tsujikawa, ``{Cosmological data
  favor Galileon ghost condensate over $\Lambda$CDM},'' {\em Phys. Rev. D} {\bf
  100} (2019), no.~6 063540, \href{http://xxx.lanl.gov/abs/1905.05166}{{\tt
  1905.05166}}.

\bibitem{kgbgood2}
N.~Frusciante, S.~Peirone, L.~Atayde, and A.~De~Felice, ``{Phenomenology of the
  generalized cubic covariant Galileon model and cosmological bounds},'' {\em
  Phys. Rev. D} {\bf 101} (2020), no.~6 064001,
  \href{http://xxx.lanl.gov/abs/1912.07586}{{\tt 1912.07586}}.

\bibitem{cham1}
J.~Khoury and A.~Weltman, ``{Chameleon fields: Awaiting surprises for tests of
  gravity in space},'' {\em Phys. Rev. Lett.} {\bf 93} (2004) 171104,
  \href{http://xxx.lanl.gov/abs/astro-ph/0309300}{{\tt astro-ph/0309300}}.

\bibitem{cham2}
J.~Khoury and A.~Weltman, ``{Chameleon cosmology},'' {\em Phys. Rev. D} {\bf
  69} (2004) 044026, \href{http://xxx.lanl.gov/abs/astro-ph/0309411}{{\tt
  astro-ph/0309411}}.

\bibitem{chamno}
J.~Wang, L.~Hui, and J.~Khoury, ``{No-Go Theorems for Generalized Chameleon
  Field Theories},'' {\em Phys. Rev. Lett.} {\bf 109} (2012) 241301,
  \href{http://xxx.lanl.gov/abs/1208.4612}{{\tt 1208.4612}}.

\bibitem{swamp1}
G.~Obied, H.~Ooguri, L.~Spodyneiko, and C.~Vafa, ``{De Sitter Space and the
  Swampland},'' \href{http://xxx.lanl.gov/abs/1806.08362}{{\tt 1806.08362}}.

\bibitem{swamp2}
H.~Ooguri, E.~Palti, G.~Shiu, and C.~Vafa, ``{Distance and de Sitter
  Conjectures on the Swampland},'' {\em Phys. Lett. B} {\bf 788} (2019)
  180--184, \href{http://xxx.lanl.gov/abs/1810.05506}{{\tt 1810.05506}}.

\bibitem{swampcos}
P.~Agrawal, G.~Obied, P.~J. Steinhardt, and C.~Vafa, ``{On the Cosmological
  Implications of the String Swampland},'' {\em Phys. Lett. B} {\bf 784} (2018)
  271--276, \href{http://xxx.lanl.gov/abs/1806.09718}{{\tt 1806.09718}}.

\bibitem{Brax:1999gp}
P.~Brax and J.~Martin, ``{Quintessence and supergravity},'' {\em Phys. Lett. B}
  {\bf 468} (1999) 40--45, \href{http://xxx.lanl.gov/abs/astro-ph/9905040}{{\tt
  astro-ph/9905040}}.

\bibitem{Brax:1999yv}
P.~Brax and J.~Martin, ``{The Robustness of quintessence},'' {\em Phys. Rev. D}
  {\bf 61} (2000) 103502, \href{http://xxx.lanl.gov/abs/astro-ph/9912046}{{\tt
  astro-ph/9912046}}.

\bibitem{Copeland:2000vh}
E.~J. Copeland, N.~Nunes, and F.~Rosati, ``{Quintessence models in
  supergravity},'' {\em Phys. Rev. D} {\bf 62} (2000) 123503,
  \href{http://xxx.lanl.gov/abs/hep-ph/0005222}{{\tt hep-ph/0005222}}.

\bibitem{cesc}
L.~Bordin, F.~Cunillera, A.~Lehebel, and A.~Padilla, ``{A natural theory of
  dark energy},'' {\em Phys. Rev. D} {\bf 101} (2020) 085012,
  \href{http://xxx.lanl.gov/abs/1912.04905}{{\tt 1912.04905}}.

\bibitem{stuck}
S.~Nagy, A.~Padilla, and I.~Zavala, ``{The Super-Stuckelberg procedure and dS
  in Pure Supergravity},'' \href{http://xxx.lanl.gov/abs/1910.14349}{{\tt
  1910.14349}}.

\end{thebibliography}\endgroup

\end{document}